\documentclass[journal,10pt]{IEEEtran}
\usepackage[utf8]{inputenc}

\usepackage{booktabs}
\usepackage{cite}
\usepackage{amsmath,amssymb,amsfonts,braket}
\usepackage{algpseudocode}
\usepackage{algorithm}
\usepackage[final]{graphicx}
\usepackage{textcomp}
\usepackage{xcolor}
\usepackage{physics}
\usepackage{amssymb}
\usepackage{bbm}
\usepackage{amsthm}
\newtheorem{proposition}{Proposition}
\usepackage{cuted}
\usepackage{capt-of}
\usepackage[colorlinks,linkcolor=blue,anchorcolor=blue,citecolor=blue,urlcolor=black]{hyperref}

\begin{document}
\hyphenation{op-tical net-works semi-conduc-tor}

\title{Communicating Properties of Quantum States over Classical Noisy Channels}

\author{Nikhitha Nunavath, Jiechen Chen, Osvaldo Simeone, Riccardo Bassoli, and Frank H. P. Fitzek 
\thanks{N. Nunavath, R. Bassoli, and F. Fitzek are with Deutsche Telekom Chair of Communication Networks, Technische Universität Dresden, Germany (email:\{nikhitha.nunavath, riccardo.bassoli, frank.fitzek\}@tu-dresden.de). R. Bassoli, and F. Fitzek are also with Centre for Tactile Internet with Human-in-the-Loop (CeTI), Dresden, Germany. J. Chen is with King’s College London, London, WC2R 2LS, UK (email:jiechen.chen@kcl.ac.uk). O. Simeone is with the Institute for Intelligent Networked Systems, Northeastern University London, One Portsoken Street, London, E1 8PH, UK (email: o.simeone@northeastern.edu).}
\thanks{The authors N. Nunavath, R. Bassoli, and F. H.P. Fitzek acknowledge the financial support by the Federal Ministry of Education and Research of Germany in the programme of “Souverän. Digital. Vernetzt.”. Joint project 6G-life, project identification number: 16KISK001K. This work is also partially funded by the German Research Foundation (DFG, Deutsche Forschungsgemeinschaft) as part of Germany’s Excellence Strategy – EXC 2050/1 – Project ID 390696704 – Cluster of Excellence “Centre for Tactile Internet with Human-in-the-Loop” (CeTI) of Technische Universität Dresden. The authors also acknowledge the financial support by the Federal Ministry of Research, Technology and Space of Germany in the project QUARKS, project identification number: 16KIS1998K. The work of O. Simeone was supported by the European Research Council (ERC) under the European Union’s Horizon Europe Programme (grant agreement No. 101198347) and by an Open Fellowship of the EPSRC (EP/W024101/1). The work by O. Simeone and J. Chen was supported also  by the EPSRC project  EP/X011852/1.}
\vspace{-1cm}
}

\maketitle

\IEEEpubidadjcol

\begin{abstract}
Transmitting information about quantum states over classical noisy channels is an important problem with applications to science, computing, and sensing. This task, however,   poses fundamental challenges due to the exponential scaling of state space with system size. We introduce shadow tomography-based transmission with unequal error protection (STT-UEP), a novel communication protocol that enables efficient transmission of properties of quantum states,  allowing decoder-side estimation of arbitrary local Pauli observables. Unlike conventional approaches requiring the transmission of a number of  bits that is exponential in the number of qubits, STT-UEP achieves communication complexity that scales logarithmically with the number of observables, depending on the observable weight. The protocol exploits classical shadow tomography for measurement efficiency, and applies unequal error protection by encoding measurement bases with stronger channel codes than measurement outcomes. We provide theoretical guarantees on estimation accuracy as a function of the bit error probability of the classical channel, and validate the approach against several benchmarks via numerical results.
\end{abstract}

\section{Introduction}

Quantum states are the foundational carriers of information in quantum technologies, from quantum computing to quantum sensing \cite{os}. They also provide a powerful mathematical framework for representing cognitive information that exhibits semantic phenomena difficult to capture with classical probability models \cite{busemeyer2012quantum}. The ability to communicate quantum states efficiently is thus essential for settings such as distributed quantum computing and sensing systems\cite{popovski20251q}, as well as  collaborative decision-making frameworks \cite{chehimi2022quantum}.

\begin{figure*}   
    \noindent 
    \makebox[\textwidth]{\includegraphics[scale=0.8]{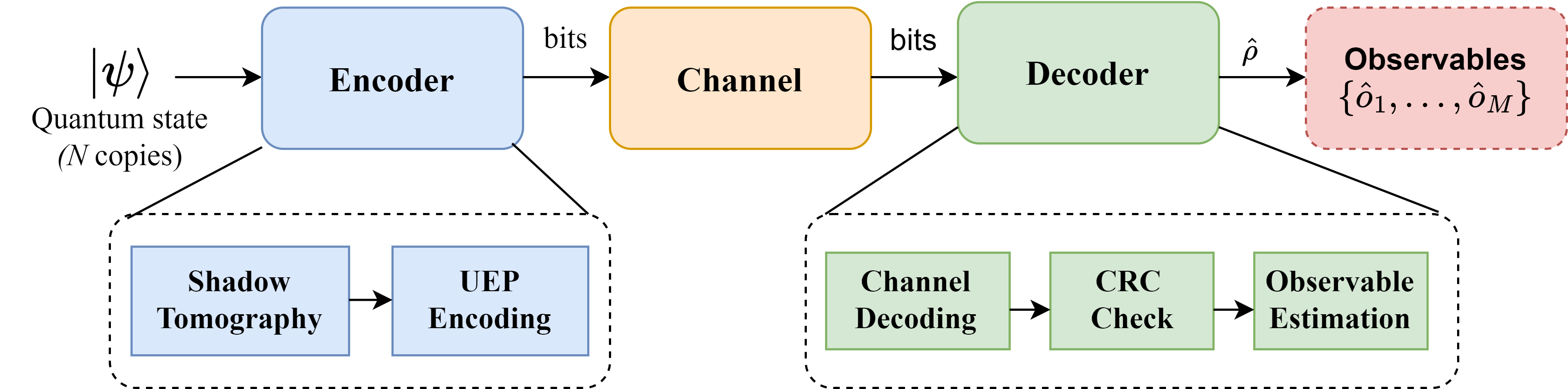}}
    \captionof{figure}{
    (top) Quantum semantic communication system: The encoder has access to $N$ copies of a quantum state $\ket{\psi}$, and aims to communicate over a noisy classical channel so that the receiver can estimate the expected values $\{\hat{o}_1, \ldots,\hat{o}_M\}$ of $M$ observables, which are a priori unknown to the transmitter. (bottom) The proposed shadow tomography-based transmission with unequal error protection (STT-UEP) applies shadow tomography to obtain observable-agnostic information that is conveyed via unequal error protection (UEP) coding to the receiver.}
    \label{fig:qst-model1}
    \vspace{-0.5cm}
\end{figure*}

However, transmitting quantum states over classical channels in the absence of pre-shared entanglement, as illustrated in Fig. \ref{fig:qst-model1}, faces fundamental challenges. Conventional approaches based on full state tomography require measurements and communication resources that scale exponentially with system size -- specifically, as  $\mathcal{O}(2^n)$ for an $n$-qubit pure state \cite{nielsen2010quantum}. This exponential scaling renders full tomography impractical for even moderately-sized quantum systems. 

However, in many communication scenarios, the receiver does not require complete state information, but rather wishes to estimate specific properties of interest, such as entanglement witnesses and fidelities with target states \cite{aaronson2018shadow}, \cite{huang2020predicting}, \cite{elben2023randomized}. As a notable example, if only bounded-weight local Pauli observables are needed, the communication complexity can be made to scale logarithmically with the number of observables, rather than exponentially with system size \cite{huang2020predicting}, \cite{elben2023randomized}. This paper leverages the semantic aspect, focusing on conveying task-relevant information instead of targeting full state reconstruction.

The work  \cite{chehimi2022quantum} introduced quantum semantic communications   as a framework in which classical information is processed via embedding on a quantum state.  Extensions of quantum semantic communications to knowledge graph transmission using entangled qubits between the encoder and the decoder were discussed in \cite{nunavath2024quantum}.

This paper introduces  shadow tomography-based transmission with unequal error protection  (STT-UEP), a communication protocol for transmitting properties of quantum  states over classical binary noisy channels (see Fig. \ref{fig:qst-model1}). We specifically aim at ensuring that the receiver can reconstruct the expected values of  an arbitrary set of $M$ observables. We make the following contributions:

\noindent $\bullet$ \textbf{Shadow tomography-based transmission with unequal error protection:} We develop STT-UEP, a transmission scheme that applies shadow tomography at the encoder via classical shadows \cite{huang2020predicting}. STT-UEP is based on the key observation that  errors in the random measurement bases are catastrophic, while errors in measurement outcomes introduce statistical noise, which is partially correctable through debiasing. Accordingly, STT-UEP applies unequal error protection by encoding bases with a smaller rate  than the outcomes, with lower rates providing  stronger protection. The encoding strategy is agnostic to the set of $M$ (weight-constrained) observables chosen at the receiver. 

\noindent $\bullet$ \textbf{Observable-independent communication complexity:} Unlike conventional approaches requiring $\mathcal{O}(2^n)$ bits for $n$-qubit pure states, STT-UEP inherits from classical shadows the benefit of requiring a number of bits that scales logarithmically with the number of observables, independently of the system size, and exponentially only in the maximum weight of the observables to be reconstructed at the receiver. We provide a theoretical result on the number of required bits as a function of the probability of error of the classical binary channel.

\noindent $\bullet$ \textbf{Experimental results:} We compare STT-UEP against the conventional quantization of state vectors  and shadow tomography with conventional coding (treating bases and outcomes equally). We demonstrate conditions under which STT-UEP achieves superior performance through its tailored unequal error protection strategy.

The remainder of this letter is organized as follows. Section II describes the system model and problem formulation. Section III presents the STT-UEP protocol and theoretical analysis. Section IV provides numerical results. Section V concludes the paper.

\begin{table}[t]
\centering
\caption{Main notation}
\label{table-I}
\begin{tabular}{ll}
\toprule
\textbf{Symbol} & \textbf{Definition} \\
\midrule
$n$ & Number of qubits \\
$N$ & Number of state copies \\
$M$ & Number of observables \\
$w$ & Maximum observable weight \\
$\varepsilon$ & Accuracy threshold \\
$\delta$ & Failure probability \\
$B$ & Number of transmitted bits \\
$R_b / R_u$ & Code rates for measurement outcomes/ measurement bases \\
\bottomrule
\end{tabular}
\vspace{-0.55cm}
\end{table}

\section{Setting and Problem Formulation}\label{Sec-II}

As illustrated in Fig. \ref{fig:qst-model1}, we consider communicating a quantum state $\ket{\psi}$ to a receiver over a noisy classical channel. The objective is to enable the receiver to estimate $M$ observables of the state. The $M$ observables are unknown to the encoder a priori, and can be arbitrarily chosen by the decoder.

\vspace{-0.3cm}

\subsection{Communication Setting}
As seen in Fig.~\ref{fig:qst-model1}, the considered  quantum semantic communication system has the following three components.

\noindent \textbf{Encoder:} The encoder has access to $N$ copies of an $n$-qubit quantum state $|\psi\rangle$. The encoder performs measurements on these copies, obtaining $Nn$ bits. These bits can be encoded, producing a packet of $B \geq Nn$ bits.

\noindent \textbf{Channel:} The $B$ encoded bits are sent over a classical channel to the decoder. 

\noindent \textbf{Decoder:} Using the $B$ classical bits received from the channel, the decoder produces estimates $\{\hat{o}_1, \ldots,\hat{o}_M\}$ for the expected values $\{\langle O_1 \rangle, \ldots, \langle O_M \rangle\}$ of $M$ observables $O_1, \ldots, O_M$. Recall that the expected value of an observable $O_m$, $m=1,\ldots,M$, for a pure state $\ket{\psi}$ is defined as \cite{os}
\begin{align}
    \langle O_m\rangle = \langle \psi|O_m| \psi \rangle. \label{OM}
\end{align}

We focus on the common case in which the observables are local Pauli observables with a maximum weight $w$. A Pauli observable $O_m$ with weight no larger than $w$ is a Hermitian operator on $n$ qubits of the form
\begin{equation}\label{Eqn-Pauli}
O_m = O_{m,1} \otimes O_{m,2} \otimes \cdots \otimes O_{m,n},
\end{equation}
where $\otimes$ denotes the Kronecker product (see, e.g., \cite{os}), and $O_{m,i} \in \{I, X, Y, Z\}$ is a Pauli matrix acting on qubit $i$.
The weight of the observable is the number of indices $i$ for which $O_{m,i}$ is different from the identity matrix $I$. For instance, a weight-2 observable measures correlations between two qubits. 
The support of each observable $O_m$ is the set of indices 
\begin{align}
    \mathcal{S}_m = \{i\in \{1,\ldots,n\} : O_{m,i} \neq I\}
\end{align}
in which the Pauli matrices are different from the identity. As mentioned, we impose the weight constraint $|\mathcal{S}_m| \leq w$, where $|\mathcal{S}_m|$ is the cardinality of set $\mathcal{S}_m$.

The goal is to ensure that the estimates $\hat{o}_1, \ldots,\hat{o}_M$ are $\varepsilon$-accurate representations of the respective true expected values $\langle O_1 \rangle, \ldots, \langle O_M \rangle$. Specifically, we impose the probabilistic requirement
\begin{equation}
   \text{Pr}(|\hat{o}_m - \langle O_m \rangle| \leq \varepsilon \text{ for all } m \in \{1, \dots M\}) \geq 1-\delta  \label{P_succ}
\end{equation}
for user-defined thresholds $\varepsilon$ and $\delta$. The requirement \eqref{P_succ} must hold for any set $M$ of local Pauli observables $O_1, \dots, O_M$, as long as the maximum weight does not exceed $w$. Note, in particular, that the encoder need not know the specific observables \eqref{Eqn-Pauli} chosen by the decoder.

\section{Shadow Tomography-Based Communication}\label{Sec-III}
In this section, we introduce shadow tomography-based transmission with unequal error protection (STT-UEP), a new communication protocol based on shadow tomography \cite{aaronson2018shadow}, \cite{huang2020predicting, nguyen2023shadow} and unequal error protection (UEP) \cite{borade2009unequal}.

\subsection{Encoder}
In STT-UEP, the encoder applies shadow tomography \cite{aaronson2018shadow}, followed by an UEP scheme tailored to the transmission of classical shadows \cite{huang2020predicting, elben2023randomized}.  Accordingly, the encoder first selects a set of unitary transformations $\{U_l\}_{l=1}^L$, which are used to pre-process each copy of the quantum state prior to a measurement in the computational basis. This pre-processing step effectively changes the measurement basis. 

For each unitary transformation $U_l=U_{l,1} \otimes \ldots \otimes U_{l,n}$, we restrict each local unitary $U_{l,i}$ to the subset $\{I, H, HS^{\dagger}\}$ of the single-qubit Clifford group, where $H$ is the Hadamard gate and $S$ the phase gate. As shown in Table I, this choice corresponds to measuring each qubit in one of the Pauli bases $\{Z, X, Y\}$, respectively. Each unitary $U_{l,i}$ is chosen independently and uniformly.

For each $i$-th copy of quantum state $\ket{\psi}$, the encoder first chooses a unitary $U_i$ uniformly at random from the described ensemble $\{U_l\}_{l=1}^L$, and then apply it to the $i$-th copy of the quantum state, resulting in the transformed state
\begin{align}\label{Eqn-5}
    \ket{\phi_i} = U_i \ket{\psi}.
\end{align}
The encoder then measures the transformed quantum state $\ket{\phi_i}$ in the computational basis, obtaining the bits $\boldsymbol{b}_i = (b_{i,1}\ldots, b_{i,n})$ with probability
\begin{align}
\Pr(\boldsymbol{b}_i|U_i|\psi\rangle) = |\langle \boldsymbol{b}_i|U_i\ket{\psi}|^2,
\end{align}
where $|\boldsymbol{b}_i\rangle = \ket{b_{i,1}}\otimes \ket{b_{i,2}} \otimes \dots\ket{b_{i,n}}$ and $b_{i,j} \in \{0, 1\}$ for all $j = 1, \dots, n$.

As discussed, measuring the transformed state $\ket{\phi_i}$ in the computational basis is equivalent to measuring the original state $\ket{\psi}$ in the Pauli basis
\begin{align}
    P_{i,j} = U_{i,j} Z U_{i,j}^{\dagger}, \label{Wij}
\end{align}
for each qubit $j$ (see Table II). In particular, each bit outcome $b_{i,j} \in \{-1, +1\}$ corresponds to an eigenvalue of the observable $P_{i,j}$, and the full measurement outcome $\boldsymbol{b}_i$ corresponds to an eigenstate of the Pauli string
\begin{equation}
    P_i = P_{i,1} \otimes P_{i,2} \otimes \ldots \otimes P_{i,n}. \label{Wi}
\end{equation}

\begin{table}[t]
\centering
\caption{Correspondence between single-qubit unitary transformations and Pauli observables.}
\label{table-II}
\begin{tabular}{cc}
\toprule
\textbf{Unitary $U$} & \textbf{Observable $U Z U^{\dagger}$} \\
\midrule
$I$  & $Z$ \\
$H$  & $X$ \\
$SH$ & $Y$ \\
\bottomrule
\end{tabular}
\vspace{-0.6cm}
\end{table}

The unitary matrices $U_i$ and the measurement outcomes $\boldsymbol{b}_i$ for the $N$ copies of quantum states, i.e., $\{(U_{1}, \boldsymbol{b}_{1}), \dots, (U_{N}, \boldsymbol{b}_{N})\}$, are represented by 
\begin{align}
    (\lceil\log_2(L)\rceil+n)N~ \text{bits},
\end{align}
where $\lceil \cdot \rceil$ is the ceiling function.  In fact, each pair $(U_i, \boldsymbol{b}_i)$ requires $\lceil\log_2(L)\rceil$ bits to identify the unitary $U_i$ within the set $\{U_l\}_{l=1}^L$, while the measurement outputs $\boldsymbol{b}_i$ amount to $n$ bits. 

STT-UEP treats the two sets of bits, describing the selected bases $\{U_1, \ldots, U_N\}$ and the measured bits $\{\boldsymbol{b}_1, \ldots, \boldsymbol{b}_N\}$, in a different way. The rationale for this choice is that, as it will be seen in this section, errors on the measurement bits can be partially mitigated via post-processing, while this is not the case for errors on the measurement bases.

To elaborate, fix a family of channel codes with rates $0<R\leq1$ and inputs given by $m$ information bits. Each such code returns $m/R$ encoded bits. For the given channel, we write the block error rate (BLER) and the bit error rate (BER) of the code with input length $m$ and rate $R$ as 
\begin{equation}\label{Eqn-BLER,BER}
    \text{BLER}(m, R) \text{ and } \text{BER}(m, R),
\end{equation} 
respectively. Both probabilities $\text{BLER}(m, R)$ and $\text{BER}(m, R)$ in \eqref{Eqn-BLER,BER} are assumed to be increasing with rate $R$, so that a lower rate $R$ yields more protection against channel errors.

STT-UEP applies a channel code of rate $0<R_b \leq1$ for the measurement bits, while the bits representing measurement bases are encoded with a more powerful channel code of rate $0<R_u<R_b$. By choosing a lower code rate for the second class of bits, we ensure that they are more protected from channel errors. This yields 
\begin{equation}\label{Eqn-B}
    B = \bigg(\frac{n}{R_b}+\frac{\lceil\log_2(L)\rceil}{R_u}\bigg)N
\end{equation}
encoded bits. 

These bits are transmitted to the receiver along with a short cyclic redundancy check (CRC) field applied to the encoded bits representing the measurement bases. The CRC field increases the number of bits $B$ in \eqref{Eqn-B} by a negligible amount, and its overhead is not added explicitly in \eqref{Eqn-B}. The addition of a CRC, which is standard in communication systems (see, e.g., \cite{cox2025introduction}), allows the detection of decoding errors for the bits encoding bases.

\subsection{Decoder}
As discussed in Sec. \ref{Sec-II}, the decoder is interested in estimating the expected values of $M$ observables $O_1, \ldots, O_M$ of the form \eqref{Eqn-Pauli}.
To this end, the decoder first applies channel decoding to recover an estimate of the $\lceil \log_2(L) \rceil N$ bits representing the measurement bases $\{U_{1}, \ldots, U_{N}\}$, as well as an estimate of the bits $\{ \boldsymbol{b}_1, \ldots, \boldsymbol{b}_N\} $. 

By using the CRC, the decoder then verifies whether the decoded bits for the measurement bases are corrupted. In particular, the CRC test reveals whether any bit errors remain uncorrected after channel decoding with a probability of incorrect error detection that decreases exponentially with the CRC size. The decoder then declares an outage if the CRC test fails, and it requests retransmission. This yields the outage probability
\begin{equation}\label{Eqn-Poutage}
P_{\text{outage}} = \text{BLER}(\lceil\log_2(L)\rceil N, R_u).
\end{equation}

If the CRC test succeeds, the decoder assumes that the bases $\{U_1, \dots, U_N\}$ are correctly decoded and continues to the next step. In the next step, after channel decoding, the measurement output bits $\{\boldsymbol{b}_1, \ldots, \boldsymbol{b}_N\}$ are received as $\{\hat{\boldsymbol{b}}_1, \ldots, \hat{\boldsymbol{b}}_N\}$, where each bit is subject to the bit flip probability 
\begin{equation}
p_{\text{err}} = \text{BER}(Nn, R_b). \label{perr}
\end{equation}
We assume without loss of generality the condition $p_{\text{err}} \leq 0.5$, and we treat the bit errors as independent across the $Nn$ bits $\{\hat{\boldsymbol{b}}_1, \ldots, \hat{\boldsymbol{b}}_N\}$. This assumption holds approximately when channel coding is preceded by interleaving \cite{shi2004interleaving}.

\begin{figure}[t]
    \centering
    \includegraphics[width=0.5\textwidth]{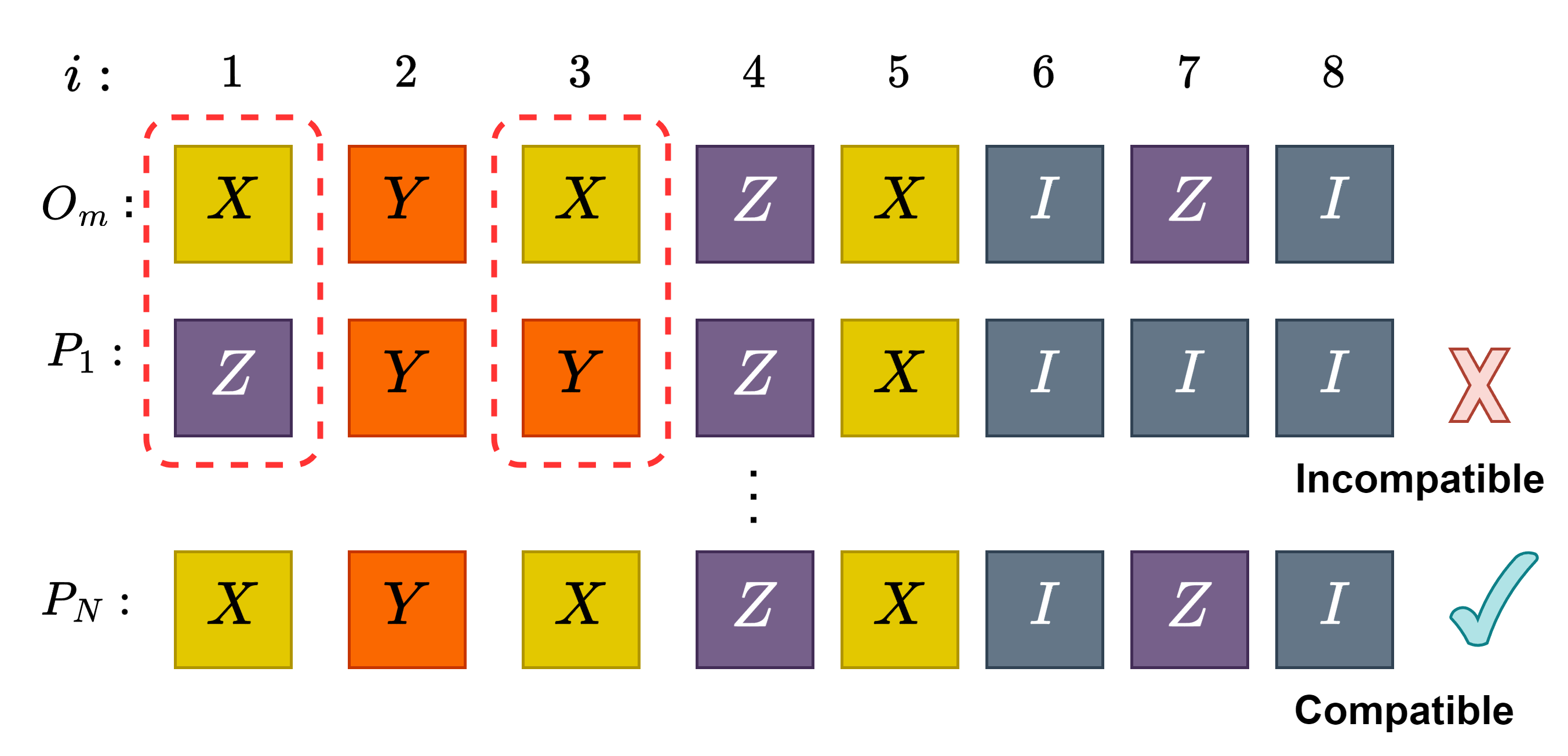}
    \vspace{-0.65cm}
    \caption{
    To estimate the expected value $\langle O_m \rangle$, the decoder retains only measurements corresponding to compatible bases $P_i$. 
    Basis $P_1$ is incompatible due to mismatches at positions 1 and 3 (highlighted with dashed red lines), while basis $P_N$ is compatible as all support positions match.}
    \label{fig:qst-decoder}
    \vspace{-0.5cm}
\end{figure}

As illustrated in Fig.~\ref{fig:qst-decoder}, based on the received unitaries $\{U_1, \dots, U_N\}$, for each observable $O_m$, the decoder selects the measurements $\hat{\boldsymbol{b}}_i$ for which the Pauli operators in the basis $P_i$ in \eqref{Wi} coincide with those of the observables $O_m$ on the support $\mathcal{S}_m$, i.e., $P_{i,j}=O_{m,j} \text{ for every } j\in\mathcal{S}_m$. We collect the indices of such measurements in the set
\begin{equation}\label{coms}
\mathcal{C}_m = \bigg\{i\in \{1,\ldots, N\}: \sum_{j \in \mathcal{S}_m} \mathbbm{1}(P_{i,j}=O_{m,j}) = |\mathcal{S}_m| \bigg\}, 
\end{equation}
where $\mathbbm{1}(\cdot)$ is the indicator function, which returns 1 if the argument is true and 0 otherwise.

For each index $i \in \mathcal{C}_m$, we have a basis $P_i$ and a received measurement outcome $\hat{\boldsymbol{b}}_i$. Evaluating product of the bits $\hat{\boldsymbol{b}}_i\in \{-1, +1\}^n$ over the support $\mathcal{S}_m$ yields
\begin{equation}
    \bar{o}_{i,m} = \prod_{j \in \mathcal{S}_m} \hat{b}_{i,j}, \label{Eqn-obar}
\end{equation}
with $\bar{o}_{i,m}\in \{-1, +1\}$.
A simple estimate of the expected value $\langle O_m \rangle$ would be the average of the outcomes \eqref{Eqn-obar} over compatible measurements, i.e., 
\begin{equation}
    \hat{o}_m^{\text{biased}} = \frac{1}{N} \sum_{i \in \mathcal{C}_m} \bar{o}_{i,m}. \label{bias}
\end{equation}
This estimator is biased due to the effect of bit flips, and due to the selection bias inherent in the basis selection procedure illustrated in Fig. \ref{fig:qst-decoder}. Specifically, bit flip errors with probability $p_{\text{err}}$ cause a systematic scaling of the expected value, while the random basis selection introduces an additional factor of $3^{-|\mathcal{S}_m|}$. However, since the probability $p_{\text{err}}$ is known for the given channel code, we can correct for this bias. As shown in the Appendix, an unbiased estimator is obtained as
\begin{align}
    \hat{o}_m = a_m \hat{o}^{\rm biased}_m, \label{om-debias}
\end{align}
where the scaling factor is
\begin{equation}
\begin{aligned}
    a_m & = \frac{3^{|\mathcal{S}_m|}}{  (1-2p_{\rm err})^{|\mathcal{S}_m|}}. \label{am}
\end{aligned}
\end{equation}
In fact, we have the equality $\mathbb{E}[\hat{o}_{m}]=a_m \mathbb{E}[\hat{o}^{\rm biased}_m] =\langle O_m \rangle$.

\subsection{Theoretical Properties of STT-UEP}
The following proposition proves that STT-UEP can guarantee the condition \eqref{P_succ} as long as the number $N$ of copies of the quantum state $|\psi\rangle$ is sufficiently large (see the Appendix for a proof).
\begin{proposition}
Consider any $M$ Pauli string observables as in \eqref{Eqn-Pauli} with maximum weight $w$, with probability $1-P_{\text{outage}}$, with $P_{\text{outage}}$ in \eqref{Eqn-Poutage}, STT-UEP guarantees the requirement \eqref{P_succ} if the number of state copies meets the inequality
\begin{equation}
N \geq \frac{2\cdot 9^{w} \ln(2M/\delta)}{(1-2p_{\text{\rm err}})^{2w}\cdot \varepsilon^2}, \label{bound}
\end{equation}
where $p_{\rm err}$ is the bit error probability  in \eqref{perr}.
\end{proposition}

\section{Experimental Results}
In this section, we evaluate the performance of the proposed STT-UEP scheme through numerical simulations and compare it against two baseline methods. 

\noindent \textbf{Benchmark Schemes:}
We consider the following benchmarks.

\noindent 1) \textbf{Direct quantization of observables (DQO):} DQO is an oracle baseline solution in which the transmitter has knowledge of $M$ observables, and allocates $N/M$ copies per observable, estimating $\langle O_m\rangle$ through direct measurement, and quantizing the $M$ estimated expected values using $b$ bits. This yields $B = M · b/R$ transmitted bits.

\noindent 2) \textbf{Conventional quantization of classical representation (CQCR):} CQCR  applies scalar quantization to the $2^n$ elements of the state vector $\ket{\psi}$, transmitting the quantized states to the receiver. Note that this is an idealistic scenario, as we only assume access to copies of the state for STT-UEP. Using a quantizer with resolution of $b$ bits and a channel code with rate $R$, this yields the total number of  transmitted bits $B = 2^{n} \cdot b/R$. Given the quantized state, say $|\hat{\psi}\rangle$, the receiver estimates the expected value of the desired observables $O_m$ as $\langle \hat{\psi} |O_m| \hat{\psi}\rangle $.

\noindent 3) \textbf{Shadow tomography-based transmission with conventional coding (STT-CC):} STT-CC is a special case of the proposed ST-UEP with the same code rate $R$ applied for both measurement bits and bases, i.e., $R_b=R_u$.

\noindent \textbf{Experimental Setup:}
The state $|\psi\rangle$ is generated independently as a Haar pure random state on $n =20$ qubits. We consider estimating the expected value of $M=30$ Pauli observables. The support positions and Pauli types are sampled uniformly at random.  We implement an LDPC code via Sionna \cite{hoydis2022sionna}, and communication takes place over an additive white Gaussian noise channel.  We set $\varepsilon = 0.2$, and we report the probability of success $P_{\rm succ}= \text{Pr}(|\hat{o}_m - \langle O_m \rangle| \leq \varepsilon \quad\text{for all } m )$.
We run the experiments over 1000 Monte-Carlo trials.

\noindent \textbf{Results and Discussion:}
\begin{figure}
    \centering
    \includegraphics[width=0.4\textwidth]{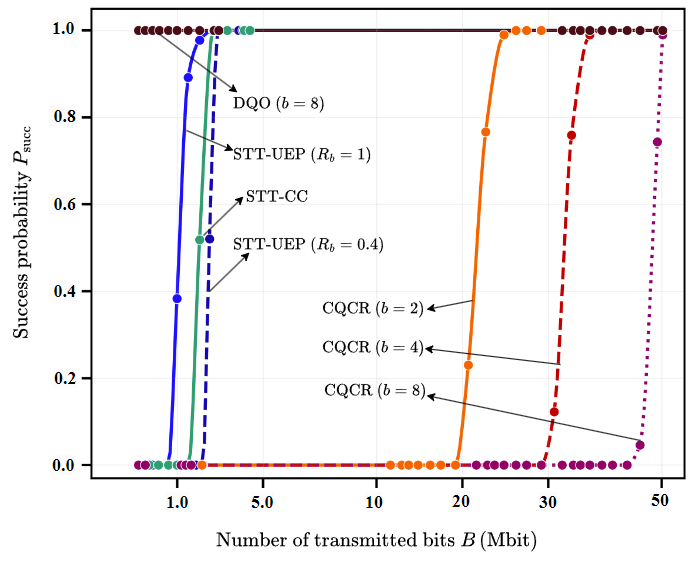}
    \vspace{-0.25cm}
    \caption{Success probability $P_{\rm succ}$ versus the number of transmitted bit $B$ for STT-UEP with code rates $R_b=0.4$ and $R_b=1$, STT-CC, DQO with quantization resolutions $b=8$ bits, and CQCR with quantization resolutions $b=2$ bits, $b=4$ bits and $b=8$ bits. }
    \label{fig:results1}
\end{figure}
Fig. \ref{fig:results1} plots the success probability $P_{\rm succ}$ as a function of the total number of transmitted bits $B$ for STT-UEP with code rates $R_b=0.4$ and $R_b=1$, STT-CC, DQO with quantization resolutions $b=8$ bits, and CQCR with quantization resolutions $b=2$ bits, $b=4$ bits and $b=8$ bits. For a fair comparison, all schemes are constrained to use the same number of transmitted bits $B$, which is achieved by appropriately adjusting the code rate $R_u$ for STT-UEP, and the code rates $R$ for STT-CC, DQO and CQCR. 

Among all other observable-agnostic schemes, the DQO baseline achieves the best performance by relying on knowledge of the observables at the transmitter. STT-UEP with uncoded transmission of measurement bits ($R_b=1$) consistently achieves the best success probability using the fewest transmitted bits $B$. In contrast, STT-CC requires more bits to reach the same success probability, since it allocates equal protection to both bases and outcomes. The CQCR baselines are the least bit-efficient, despite requiring knowledge of the classical description of the state.

\begin{figure}
    \centering
    \includegraphics[width=0.4\textwidth]{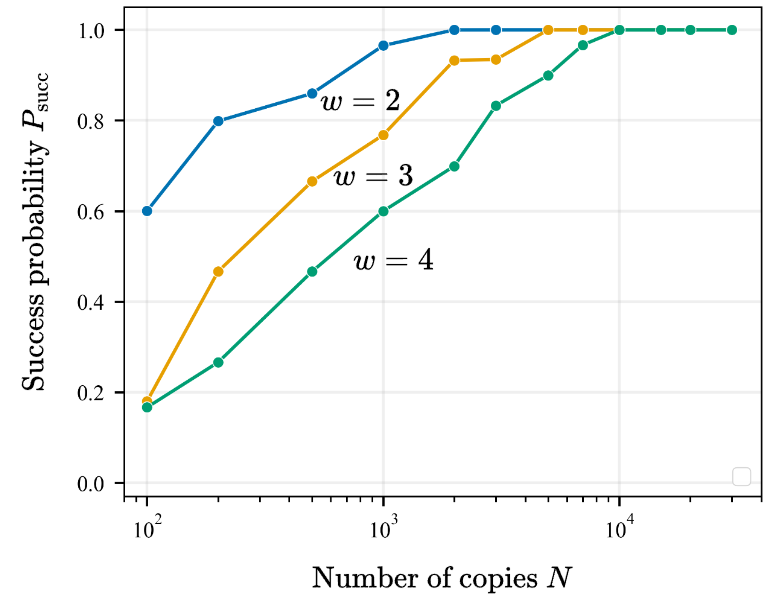}
    \vspace{-0.25cm}
    \caption{Success probability $P_{\rm succ}$ versus the number of copies $N$ for STT-UEP with $R_u=0.4$ and $R_b=1$.}
    \label{fig:results2}
    \vspace{-0.5cm}
\end{figure}

Fig.~\ref{fig:results2} shows the success probability $P_{\rm succ}$ as a function of the number of copies $N$ for the proposed STT-UEP scheme with code rates $R_u=0.4$ and $R_b=1$ for Pauli observables with weights $w\in\{2,3,4\}$. Confirming insights from Proposition 2, the success probability increases with the number of copies $N$, and observables with higher weight require a larger number of copies to achieve the same success probability.

\section{Conclusions}
This work has studied the problem of communicating properties of a quantum state over a classical noisy channel, with the goal of enabling the recovery of arbitrary local Pauli observables at an decoder. Future research directions include extending the framework to fading channels, developing adaptive coding strategies, exploring adaptive measurement schemes with feed-forward basis adjustment, and exploring multi-user or distributed quantum sensing scenarios.

\bibliographystyle{IEEEtran}
\bibliography{name}

\section*{Appendix }
To prove the unbiasedness of \eqref{om-debias}, note that the $i$-th basis $P_i$ is compatible with the observable $O_m$ with probability  $3^{-|\mathcal{S}_m|}$. Furthermore, since the channel errors are assumed to be independent with probability $p_{\rm err}$, the probability that even number of bits are flipped is given by 
\begin{align}
    p_{\rm even} = \frac{1+(1-2p_{\rm err})^{|\mathcal{S}_m|}}{2}. \label{peven}
\end{align}
As a result, the expectation $\mathbb{E}\big[\prod_{j\in\mathcal{S}_m} \hat{b}_{i,j} \big]$ is given by
\begin{align}
    \mathbb{E}\bigg[\prod_{j\in\mathcal{S}_m} \hat{b}_{i,j} \bigg]  = (1-2p_{\rm err})^{|\mathcal{S}_m|} \cdot \langle O_m \rangle \label{eq33}
\end{align}
yielding the desired result 
\begin{align}
    \mathbb{E}[\hat{o}_m^{\text{biased}}] &~ = 3^{-|\mathcal{S}_m|} \cdot(1-2p_{\rm err})^{|\mathcal{S}_m|} \cdot \langle O_m \rangle.
\end{align}

To prove \eqref{bound}, we leverage Hoeffding’s inequality together with the union bound. Specifically, applying the union bound to the left-hand side of \eqref{P_succ}, to guarantee \eqref{P_succ}, it suffices to require the inequalities $\text{Pr}(|\hat{o}_m - \langle O_m \rangle| > \varepsilon ) < \delta/M$ for all $m=1,\ldots, M$. By \eqref{om-debias}, the estimate $\hat{o}_m$ is supported within the interval $[-a_m,a_m]$. Thus, 
applying Hoeffding’s inequality to the left-hand side of \eqref{P_succ} yields
\begin{align}
   \text{Pr}(|\hat{o}_m - \langle O_m \rangle| > \varepsilon ) < 2 \exp \bigg(- \frac{2N \epsilon^2}{(a_m-(-a_m))^2} \bigg).
\end{align}
Using \eqref{am} and considering the worst case $|\mathcal{S}_m|=w$, we finally obtain \eqref{bound}.

\end{document}